# Coupled-Mode Theory of Field Enhancement in Complex Metal Nanostructures

Gregory Sun[1]

Department of Physics, University of Massachusetts Boston

Boston, Massachusetts 02125

Jacob B. Khurgin

Department of Electrical and Computer Engineering, Johns Hopkins University

Baltimore, Maryland 21218

Alexander Bratkovsky

Hewlett Packard Labs, Palo Alto, CA 94304

**Abstract**

We describe a simple yet rigorous theoretical model capable of analytical estimation of plasmonic field enhancement in complex metal structures. We show that one can treat the complex structures as coupled multi-pole modes with highest enhancements obtained due to superposition of high order modes in small particles. The model allows one to optimize the structures for the largest possible field enhancements, which depends on the quality factor $Q$ of the metal and can be as high as $Q^2$ for two spherical particles. The "hot spot" can occur either in the nano-gaps between the particles or near the smaller particles. We trace the optimum field enhancement mechanism to the fact that the extended dipole modes of larger particles act as the efficient antennas while the modes in the gaps or near the smaller particles act as the compact sub-wavelength cavities. We also show how easily our approach can be extended to incorporate large numbers of particles in intricate arrangements.

[1] Corresponding author, email: greg.sun@umb.edu

## I. Introduction

It has been known for many years that collective oscillations of electrons in metals structured on the sub-wavelength scale are capable of exciting local optical fields that exceed the average fields impinging on the structure by orders of magnitude. This phenomenon had been successfully used to demonstrate spectacular enhancement of sensitivity in Raman sensing [1-5] as well as in fluorescence measurements [6-9], and had been proposed as a method to increase the efficiency of solar cells [10], detectors [11], and various nonlinear optical devices.

In addition to these impressive experimental results, a better picture of understanding the local field enhancement has been gradually emerging thanks to the efforts of a large community of theorists involved in the nano-plasmonics research. It has become clear that the enhancement is the largest in the so-called "hot-spots" [12-14] occurring when the metal is structured in a rather sophisticated way with sharp peaks or small gaps. The maximum enhancement is limited by the metal loss. A single metal nanoparticle [15-25] having a simple smooth shape (sphere, ellipsoid, or nano-rod) usually provides the electric field enhancement no larger than a $Q$-factor of the metal [26,27], where $Q = \varepsilon''/\varepsilon'$ is the ratio of real and imaginary parts of dielectric function of the metal, and is less than a factor of 10-20 in the visible and near IR. But far more significant (up to three orders of magnitude) enhancement can occur in the intricately structured and arranged nanoparticles when the field gradually couples from the larger particles or regions serving as antennae into the smaller regions that serve as field-concentrating hot spots [28].

It has been suggested by Stockman [29] that sequential coupling of energy from larger to smaller particles can result in high degree of field concentration. At the same time, Norlander's group [30] have pioneered the plasmon-hybridization formalism in nanoshells and their dimers [31-53] and trimers and more complex structures, where the highest field concentration is achieved in the gaps, similar to the bow-tie nano-antennas [54,55].

Yet for the most part the theoretical description of the field enhancement in the complex plasmonic nanostructures relies heavily on time-consuming numerical simulations, thus the basic physics behind the enhancement tends to become obscured making optimization rather difficult. Furthermore, as we have mentioned before, in the numerical calculations the radiative losses [56] are not always taken into account correctly, as emphasized in our prior works [57]. When it comes to analytical models, hybridization model [30] predicts the position of spectral peak rather precisely, but it does not provide analytical expressions for the field enhancement. In addition, when the damping rate becomes commensurate with the coupling terms (which is often the case) the hybrid states model fails and one has to consider the coupling and damping processes on equal footing, which, to the best of our knowledge had not been done in any analytical model.

In this work we develop a fully analytical "coupled modes model" for plasmonic optical field enhancement in complex metal nanostructures. Using the model, we show that whether the enhancement is achieved near the small feature (nanotip) or inside the nano-gap, the enhancement is proportional to $Q^m$ where $m$ is the effective number of sequential coupling transitions occurring between the light being coupled into the structure and it being concentrated around or inside the smallest surface feature. Armed with these results, we develop the optimization routine for maximum field enhancement.

## II. Field enhancement theory

The field enhancement by two coupled metal nanospheres can be formulated based on our previous description of an isolated single metal sphere with a radius $a$, whose eigen modes of index $l$ in a spherical polar coordinate system as shown in Fig. 1(a), under the electro-static approximation can be given as [58]

$$\Phi_l = \begin{cases} \dfrac{a}{l+1} E_{max,l} \left(\dfrac{r}{a}\right)^l P_l(\cos\theta), & r < a \\ \dfrac{a}{l+1} E_{max,l} C \left(\dfrac{a}{r}\right)^{l+1} P_l(\cos\theta), & r \geq a \end{cases} \quad (1)$$

where $P_l(\cos\theta)$ is the Legendre polynomial and $E_{max,l}$ is the maximum electric field located just outside of the metal sphere at $r = a$ and $\theta = 0$. The theory can be adapted rather easily to the elliptical particles of various eccentricities, but, other than the shift in the resonance frequency, the conclusions, at least qualitatively, will not change relative to the spherical particle, while the simplicity will be lost. Therefore we shall restrict ourselves to spherical particles and their combinations to present what is essentially an analytical model. Furthermore, we shall consider only the combinations of nanoparticles with axial symmetry, hence we shall consider only $m_l = 0$ eigen modes, disregarding their $2l + 1$ degeneracy.

The electric field of the $l$-th mode supported by the metal sphere that is surrounded by a medium with a dielectric constant $\varepsilon_D$ can now be written as [58]

$$E_l(r,\theta) = \begin{cases} E_{in,l} \\ E_{out,l} \end{cases}$$
$$= \begin{cases} E_{max,l} \left(\frac{r}{a}\right)^{l-1} \left[-\frac{l}{l+1} P_l(\cos\theta)\hat{r} + \frac{1}{\sin\theta}[P_{l+1}(\cos\theta) - \cos\theta P_l(\cos\theta)]\hat{\theta}\right], r < a \\ E_{max,l} \left(\frac{a}{r}\right)^{l+2} \left[P_l(\cos\theta)\hat{r} + \frac{1}{\sin\theta}[P_{l+1}(\cos\theta) - \cos\theta P_l(\cos\theta)]\hat{\theta}\right], \quad r \geq a \end{cases} \quad (2)$$

The radial dependence of the electric field shows that the mode gets "compressed" closer to the surface of a nanoparticle and the mode frequency $\omega_l = \omega_p \sqrt{\frac{l}{l+(l+1)\varepsilon_D}}$ approaches $\omega_\infty = \omega_p/\sqrt{1+\varepsilon_D}$ as the mode order increases [54], where $\omega_p$ is the metal Plasmon frequency.

The surface charge density for the $l$-the mode can be evaluated using the normal component ($\hat{r}$) of the electric field in Eq. (2) at $r = a$,

$$\sigma_l(\theta) = \varepsilon_0(\varepsilon_M - 1)E^{\hat{r}}_{in,l}(a,\theta) - \varepsilon_0(\varepsilon_D - 1)E^{\hat{r}}_{out,l}(a,\theta)$$
$$= \frac{2l+1}{l+1}\varepsilon_0 E_{max,l} P_l(\cos\theta). \quad (3)$$

where $\varepsilon_0$ is the permittivity of free space. The effective volume of the $l$-th mode [59] can be defined through the mode energy $U_l = \frac{1}{2}\oiint \Phi_l \sigma_l d^2r = \frac{1}{2}\varepsilon_0\varepsilon_D E^2_{max,l} V_{eff,l}$ to arrive at

$$V_{eff,l} = \frac{4\pi a^3}{(l+1)^2 \varepsilon_D} \quad (4)$$

which is always less than the volume of the nanosphere. As the mode order index $l$ increases, the effective volume decreases roughly with $l^{-2}$ as the SP energy gets concentrated within a narrow angle around the $z$-axis near the surface of the nanosphere.

The higher order modes are obviously desirable for achieving tremendous peak energy densities. But in order to exploit these modes, one must first be able to couple external excitation into them, and here lies the main issue with the higher order modes in a symmetric spherical particle – they are completely uncoupled from the radiation modes.

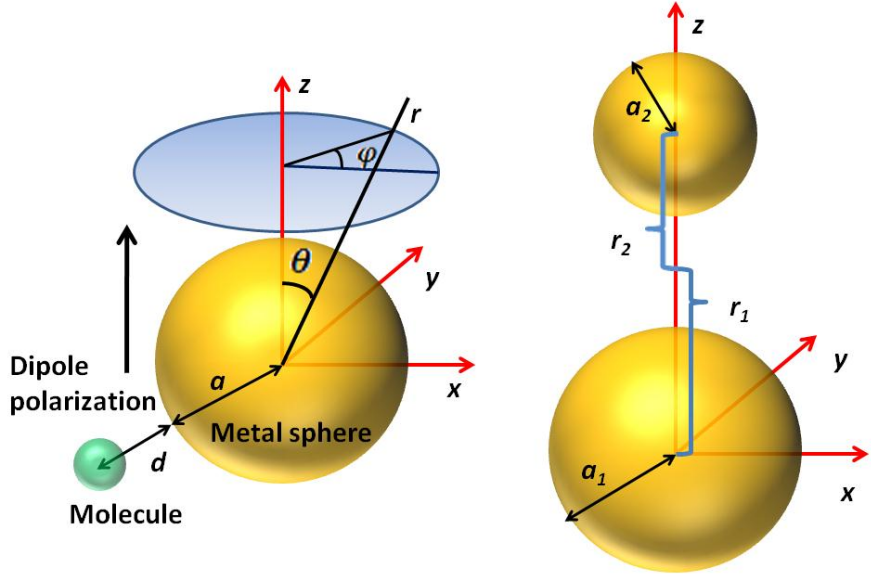

Fig. 1 Illustration of (a) the spherical coordinate system used to describe the metal sphere whose dipole is polarized along *z*-axis with a radius $a$ and (b) the geometry of two coupled metal spheres that are separated by $r_0 = r_1 + r_2$.

Indeed, the dipole moment evaluated as an integral of the charge density Eq. (3) over the sphere surface vanishes for all higher order modes ($l \geq 2$), except the $l = 1$ mode whose dipole $p_1 = 2\pi a^3 \varepsilon_0 E_{max,1}$. This dipole mode usually referred to as a localized SP mode of the nanosphere is the only solution coupled to the external fields for as long as the nanosphere diameter is much smaller than the wavelength. Therefore the dipole mode is also the only one subjected to the radiative dumping and using the standard expression for the dipole radiating power, it is easy to show that the radiative decay rate of the dipole mode [60]

$$\gamma_{rad} = -\frac{1}{U_1}\left(\frac{dU_1}{dt}\right)_{rad} = \frac{2\omega_1}{3\varepsilon_D}\left(\frac{2\pi a}{\lambda_1}\right)^3 = \frac{2\omega}{3\varepsilon_D}\chi^3 \qquad (5)$$

where $\omega_1$ is the dipole oscillating frequency, $\lambda_1$ is the corresponding wavelength in the dielectric, and $\chi$ is the normalized metal sphere radius. Simultaneously, all the modes also experience nonradiative decay due to the imaginary part of the metal dielectric function at roughly the same rate that is equal to the metal loss in the Drude model $\gamma_{nrad,l} \approx \gamma$. The decay rate can thus be summarized for all modes as

$$\gamma_l = \begin{cases} \gamma_{rad} + \gamma, & l = 1 \\ \gamma, & l \geq 2. \end{cases} \quad (6)$$

The higher order modes with smaller effective mode volumes and not subjected to radiative damping should in principle provide excellent confinement for the enhancement of optical processes. Unfortunately, these modes do not couple well into radiation modes outside the nanoparticle because of their vanishing dipole moments. The only mode that does couple to outside is the dipole mode ($l = 1$) which, on the other hand, has relatively larger effective mode volume, and thus can act as an efficient antenna but not as a good resonator. In order to achieve strong enhancement of optical properties, one needs both antenna and resonator to be efficient. But, a single mode in a symmetric structure cannot simultaneously accomplish both.

We have first encountered this challenge while analyzing single spherical nanoparticles and attempting to maximize the enhancement by optimizing the nanoparticle size. The results were far from spectacular, of course, because large particles acted as good antennae but poor resonators, and small particles vice versa. As we have already mentioned the maximum attainable field enhancement was less than $Q$ of the metal. It is only natural then to follow the techniques used in micro-wave engineering, where no one ever dreams of combining antenna and cavity into one element, but rather use two distinct elements, antenna and resonator coupled to each other.

Combining two or more nanoparticles (Fig.1(b)) allows us to engineer the schemes in which efficient antennas are coupled into the resonators with high confinement. One can think of two ways of attaining this. In case of two spheres of equal dimensions the dipole modes in both spheres act as antennae and the superposition of higher order modes act as resonators allowing efficient coupling of the radiation into the gap region. This is the dimer case considered in [30]. In case of two highly dissimilar spheres or nanolens [29] only the dipole modes participate in energy concentration. The dipole mode of the larger sphere acts as an antenna and the dipole mode of the smaller particle acts as a resonator. Obviously when the spheres are of different size yet still comparable, the field enhancement mechanism is a combination of both aforementioned effects. In this work, we explore this enhancement mechanism for a variety of nanoparticle sizes and their relative placements.

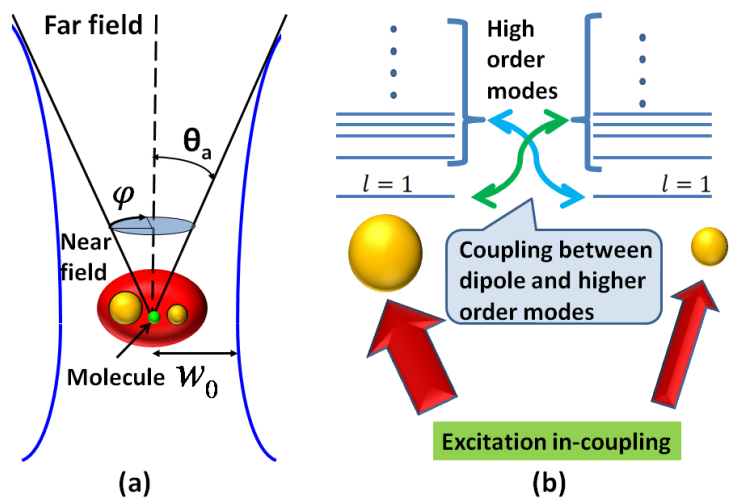

Fig. 2 Illustration of (a) the metal spheres placed at the apex of a focused Gaussian beam with a numerical aperture characterized by the far-field half angle $\theta_a$ and (b) the coupling of optical excitation into the dipole modes of both spheres and their subsequent coupling into the higher order modes.

## III. Enhancement Mechanism in Coupled Mode Theory

The way to evaluate unambiguously the field enhancement is to compare maximum field to that of tightly focused light beam in the absence of metal spheres. Consider now in the absence of metal nanoparticles an optical excitation at the frequency of $\omega$ and corresponding wavelength $\lambda$ in the dielectric medium in the form of Gaussian beam characterized by a far field half angle $\theta_a$ gets focused onto a diffraction limited spot with a radius $w = \frac{\lambda}{\pi\theta_a}$. The field in the focal spot $E_{foc}$ is related to the incident power $|s_+|^2 =$ $\frac{n}{Z_0}\pi\left(\frac{w}{2}\right)^2 E_{foc}^2$ where $Z_0$ is the impedance of free space and $n$ is the index of refraction in the dielectric [61].

We now treat the coupling of two closely spaced metal nanoparticles whose SP modes are overlapping with each other. First, we introduce the amplitude of the $l$-th mode in the $i$-th sphere using the "canonical amplitude" of the field as a square root of its energy

$$A_l^{(i)} = \sqrt{U_l^{(i)}} = \sqrt{\frac{1}{4}\varepsilon_0\varepsilon_D V_{eff,l}^{(i)}} E_{max,l}^{(i)} \tag{7}$$

Next, we obtain the coupling energy as an as an integral of the electric potential $\Phi_{l_1}^{(1)}$ of the $l_1$-th mode of sphere 1 multiplied by the surface charge density $\sigma_{l_2}^{(2)}$ of the $l_2$-th mode of sphere 2 evaluated over the surface of the sphere 2,

$$U_{l_1 l_2}^{(12)} = \oiint \Phi_{l_1}^{(1)} \sigma_{l_2}^{(2)} ds^{(2)} = -4\kappa_{l_1 l_2}^{(12)} A_{l_1}^{(1)} A_{l_2}^{(2)}. \tag{8}$$

Of all the coupling coefficients $\kappa_{l_1 l_2}^{(12)}$ we are mostly interested in the coupling between the dipole mode ($l_1 = 1$) in one sphere and all the modes ($l_2 = l$) in the other sphere $\kappa_{1l}^{(12)}$ because they are the only ones associated with energy transfer, which can be obtained analytically,

$$\kappa_{1l}^{(12)} = \frac{l+1}{2}\left(\frac{a_1}{r_0}\right)^{3/2}\left(\frac{a_2}{r_0}\right)^{l+1/2}. \tag{9}$$

The coupling between higher order modes in two spheres only shifts the resonant frequencies of those modes by a small amount, typically smaller than broadening $\gamma$ and can be neglected in this analysis.

Now the energy balance equations for sphere $i$ can be written for its dipole ($l = 1$) and higher orders ($l \geq 2$) modes separately as

$$\begin{aligned}\frac{dA_1^{(i)}}{dt} &= j(\omega_1 - \omega)A_1^{(i)} - j\sum_{l=1}^{\infty}\omega_{1l}\kappa_{1l}^{(ij)}A_l^{(j)} - \frac{1}{2}\gamma_1^{(i)}A_1^{(i)} + \kappa_{in}^{(i)}s_+,\\ \frac{dA_l^{(i)}}{dt} &= j(\omega_l - \omega)A_l^{(i)} - j\omega_{1l}\kappa_{1l}^{(ji)}A_1^{(j)} - \frac{\gamma}{2}A_l^{(i)}, \quad l \geq 2\end{aligned} \tag{10}$$

where $\omega_{1l} = \sqrt{\omega_1 \omega_l}$. Note that only the dipole mode allows the optical excitation with the incident power $|s_+|^2$ to be coupled in with an in-coupling coefficient $\kappa_{in}$ related to the dipole radiative decay rate $\gamma_{rad}$ by $\kappa_{in} \approx \frac{\theta_a}{2}\sqrt{\frac{3\gamma_{rad}}{2}}$, according to the reciprocity by Haus [61].

At steady state, Eq. (10) relates the electric field of the $l$-th higher order mode of sphere 2 to that of dipole mode of sphere 1 as

$$E_{max,l}^{(2)} = \frac{\omega_{1l}}{(\omega_l - \omega) + j\frac{\gamma}{2}} \left(\frac{l+1}{2}\right)^2 \left(\frac{a_1}{r_0}\right)^{3/2} \left(\frac{a_2}{r_0}\right)^{l+1/2} \left(\frac{a_1}{a_2}\right)^{3/2} E_{max,1}^{(1)}. \quad (11)$$

A similar expression exists between $E_{max,l}^{(1)}$ and $E_{max,1}^{(2)}$. Let us take a quick look at Eq.(11) for two extreme cases. In the first case of "symmetric dimer", we consider two spheres of equal radii $a_2 = a_1 = a$, with negligibly small gap $r_0 \approx 2a$ and neglect the detuning relative to broadening, which leads to

$$\left|E_{max,l}^{(2)}\right| < Q \frac{(l+1)^2}{2^{l+3}} \left|E_{max,1}^{(1)}\right|, \quad Q = \frac{\omega}{\gamma}, \quad (12)$$

where we have introduced the material qualify factor $Q$ which is the ratio of the real and imaginary parts of the electric permittivity of the metal. For the noble metals in the optical and near IR regions, the value of $Q$ ranges from 10 to 15 in case of Au and it can be as high as 40 for Ag, although in the nanoparticles the actual $Q$ is always lower due to the surface scattering. If we define the cut-off mode as the one whose maximum filed is equal to ½ of the field of the dipole mode, we obtain that for realistic $Q$'s of less than 20 no more than 10 modes will get excited and once one takes detuning and gap into account that number will become even less. One can perform summation of Eq.(12) to obtain the maximum enhancement relative to single sphere with zero gap

$$2\left|1 + \frac{\sum_{l=2}^{\infty} E_{max,l}^{(2)}}{E_{max,1}^{(1)}}\right| \sim 2\left[1 + \left(\frac{9Q}{8}\right)^2\right]^{1/2} \quad (13)$$

where the factor of two in front comes from having two antennae and the factor of $9Q/8$ comes from all higher order modes of the other sphere, all added in phase. One cannot help but refer to this phenomenon as "spatial mode-locking".

In the other extreme of "nanolens" $a_2 \ll a_1$ only the larger sphere would act as an antenna and for all the modes in the smaller sphere

$$\left|E^{(2)}_{max,l}\right| < Q \frac{(l+1)^2}{2} \left(\frac{a_2}{a_1}\right)^{l-1} \left|E^{(1)}_{max,l}\right|. \tag{14}$$

It immediately follows that in the "nanolens" regime only the $l = 1$ dipole mode gets excited, and the field enhancement relative to single sphere, in the limit of zero gap between the particles is about

$$\left|1 + \frac{E^{(2)}_{max,1}}{E^{(1)}_{max,1}}\right| \sim (1 + 4Q^2)^{1/2} \tag{15}$$

in general agreement with Ref. [29]. From the most simple considerations the "symmetric dimer" and "nanolens" can provide roughly the same field enhancement, but those are the extreme cases, and in order to optimize the field enhancement one should obtain the solution for arbitrary radii ratio and for finite gap size also taking into consideration

## IV.  Solution for the field enhancement

The total electric field at the location $r_1 = r_0 - r_2$ in the gap (Fig.1(b)) is the summation of all modes from both spheres as

$$\begin{aligned}E(r_1) = E^{(1)}_{max,1}&\left[\left(\frac{a_1}{r_1}\right)^3 + \sum_{l=2}^{\infty} \frac{\omega_{1l}\kappa^{(12)}_{1l}}{(\omega_l - \omega) + j\frac{\gamma}{2}} \frac{l+1}{2}\left(\frac{a_1}{r_2}\right)^{3/2}\left(\frac{a_2}{r_2}\right)^{l+1/2}\right]\\ + E^{(2)}_{max,1}&\left[\left(\frac{a_2}{r_2}\right)^3 + \sum_{l=2}^{\infty} \frac{\omega_{1l}\kappa^{(21)}_{1l}}{(\omega_l - \omega) + j\frac{\gamma}{2}} \frac{l+1}{2}\left(\frac{a_2}{r_1}\right)^{3/2}\left(\frac{a_1}{r_1}\right)^{l+1/2}\right]\end{aligned} \tag{16}$$

where the first term is the combination of the dipole mode of sphere 1 and the higher order modes of sphere 2, and the energy of all these modes is coupled in through the $l = 1$ mode of sphere 1, and vice versa for the second term. Applying Eq.(10) at steady state, we can relate

$$\begin{pmatrix}E^{(1)}_{max,1}\\E^{(2)}_{max,1}\end{pmatrix} = \frac{\omega}{\sqrt{2}} M^{-1}_{2\times 2} \begin{pmatrix}E_{foc}\\E_{foc}\end{pmatrix} \tag{17}$$

where the elements in the $2 \times 2$ matrix $M_{2\times 2}$ are

$$m_{11} = j(\omega - \omega_1) + \sum_{l=2}^{\infty} \frac{\omega_{1l}^2 \left[\kappa_{1l}^{(12)}\right]^2}{j(\omega - \omega_l) + \frac{\gamma}{2}} + \frac{1}{2}\gamma_1^{(1)}$$

$$m_{12} = j\omega_1 \kappa_{11} \left(\frac{a_2}{a_1}\right)^{3/2} = j\omega_1 \left(\frac{a_2}{r_0}\right)^3$$

$$m_{21} = j\omega_1 \kappa_{11} \left(\frac{a_1}{a_2}\right)^{3/2} = j\omega_1 \left(\frac{a_1}{r_0}\right)^3 \tag{18}$$

$$m_{22} = j(\omega - \omega_1) + \sum_{l=2}^{\infty} \frac{\omega_{1l}^2 \left[\kappa_{1l}^{(21)}\right]^2}{j(\omega - \omega_l) + \frac{\gamma}{2}} + \frac{1}{2}\gamma_1^{(2)}.$$

Finally, substituting Eq. (17) into Eq. (16) in conjunction with Eq. (9), we arrive at the enhancement factor which is defined as the ratio of the electric field in the presence of the metal spheres to that of the focusing spot in the absence of the metal spheres

$$F = \left|\frac{E(r_1)}{E_{foc}}\right| = \frac{\omega}{\sqrt{2}} \frac{1}{|M_{2\times 2}|} \times$$

$$\left|(m_{22} - m_{12})\left(\frac{a_1}{r_1}\right)^3 \left[1 + \frac{1}{4}\frac{a_2 r_1^3}{r_0^2 r_2^2} \sum_{l=2}^{\infty} \frac{\omega_{1l}(l+1)^2}{j(\omega_l - \omega) + j\frac{\gamma}{2}} \left(\frac{a_2^2}{r_0 r_2}\right)^l\right]\right. \tag{19}$$

$$\left. + (m_{11} - m_{21})\left(\frac{a_2}{r_2}\right)^3 \left[1 + \frac{1}{4}\frac{a_1 r_2^3}{r_0^2 r_1^2} \sum_{l=2}^{\infty} \frac{\omega_{1l}(l+1)^2}{j(\omega_l - \omega) + j\frac{\gamma}{2}} \left(\frac{a_1^2}{r_0 r_1}\right)^l\right]\right|$$

We shall now simplify Eq. (19) by examining the field enhancement at the mid gap position $r = r_0/2$ of two equal spheres $a = a_1 = a_2$ that are excited at the dipole mode frequency $\omega = \omega_1$. We use the fact that coupling coefficients are small, $\left[\kappa_{1l}^{(ij)}\right]^2 \approx 0$, and realize that the terms from higher order modes ($l \geq 2$) in Eq. (19) are significant only for those lower indexes $l$ whose frequency detuning from $\omega_1$ is small, we thus approximate $Q^{-1} \gg 2|1 - \omega_l/\omega_1|$. In the limit of zero gap, $r_0 \approx 2a$, we have

$$F \approx 2\sqrt{2}\, Q \left|1 - j\frac{9Q}{8}\right| \approx \frac{9\sqrt{2}}{4} Q^2. \tag{20}$$

In comparison with the field enhancement by a single metal sphere which is proportional to $Q$, we now have additional contributions from higher order modes that have a relationship of $Q^2$.

In the other extreme of "nanolens" $a_2 \ll a_1$, where only the dipole modes get excited and the field is focused in the vicinity of the smaller sphere, Eq. (19) reduces to

$$F \approx \sqrt{2}Q \left|1 - j2Q\left(\frac{a_1}{r_0}\right)^3\right| \left(\frac{a_2}{r_2}\right)^3. \tag{21}$$

Once again in the limit of zero gap, $r_2 \approx a_2$, and $r_0 \approx a_1$, we have $F \approx \sqrt{2}Q(1 + 4Q^2)^{1/2} \approx 2\sqrt{2}Q^2$. Obviously, both extremes significantly overestimate the enhancement, and once the detuning between the resonance frequencies of different order modes and the presence of the gap between the particles are taken into account, the actual enhancement will be significantly less as shown in the next session. Nevertheless, these two expressions do provide a quick estimate for the upper limit of the enhancement.

## V. Results and Discussion

We have used Eq. (19) to evaluate the enhancement by Au metal [62] spheres embedded in GaN dielectric with $Q \approx 10$ at the dipole frequency $\hbar\omega_1 = 1.967$eV. While it is not difficult to evaluate the field enhancement anywhere in the near field of the two spheres, we shall present our results in the gap of the two spheres since that is where the strongest enhancement occurs. We first calculate the enhancement at the frequency of optical excitation in resonance with the dipole frequency $\omega = \omega_1$. For gaps less than 2 nm, the quantum effects such as electron tunneling and screening significantly reduce the enhancement [63], we shall therefore limit our model to the coupled metal nanospheres with their separation gap greater than 2 nm. As has been demonstrated in our earlier work for isolated single spheres [27,57,58,64], the enhancement has a strong dependence on the nanoparticle size, the enhancement in the coupled structure here shown in Fig. 3 also depends quite sensitively on the sizes of both spheres. The results in Fig. 3 are for enhancement at the mid gap between the two spheres with two different gaps. For smaller gap (5 nm), there are two symmetrical peaks in Fig. 3(a) indicating that the maximum enhancement is obtained with two unequal metal spheres, i.e. the "nanolens" case. In this situation, the larger sphere primarily acts as an antenna for energy to be coupled into the system while the smaller one behaves like a cavity for energy to be concentrated. As the gap increases, the two peaks merge into one as shown in Fig. 3(b) (10 nm gap) calling for spheres of equal size. This is easy to understand because at large distances the larger antenna sphere cannot effectively excite the smaller sphere and it is preferable to have both spheres of equal size.

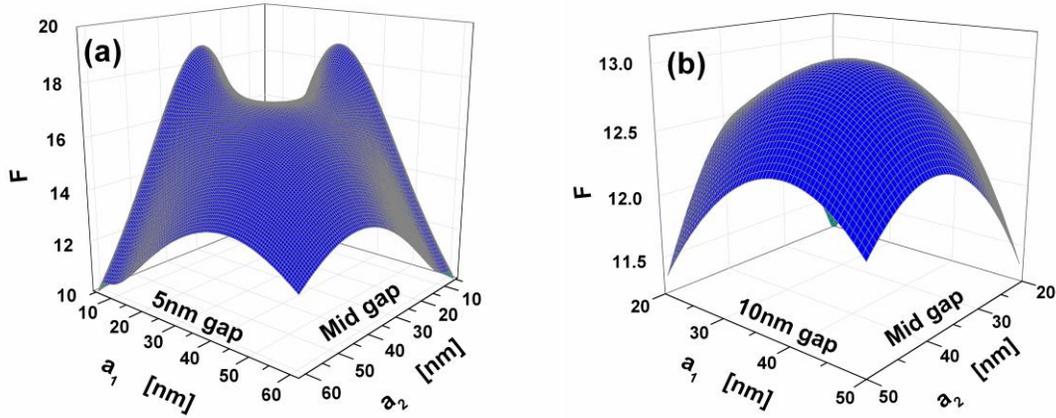

Fig.3 Enhancement $F$ at mid gap of the two Au spheres embedded in GaN as a function of their radii $a_1$ and $a_2$ with (a) 5nm and (b) 10nm gap.

Next, we calculate the enhancement at the location that is fixed at 2 nm from sphere 2 with the radius $a_2$ as shown in Fig. 4. For smaller gap of 5 nm, the location of 2 nm from sphere 2 is close to the mid gap, similarly to the result in Fig. 3(a), two peaks emerge but this time they are asymmetric, but for larger gap of 10 nm, only one peak appears. In both cases, since the position of enhancement is closer to sphere 2, it consistently favors sphere 2 to be smaller for the field to be focused in its vicinity, i.e. the "nanolens" case.

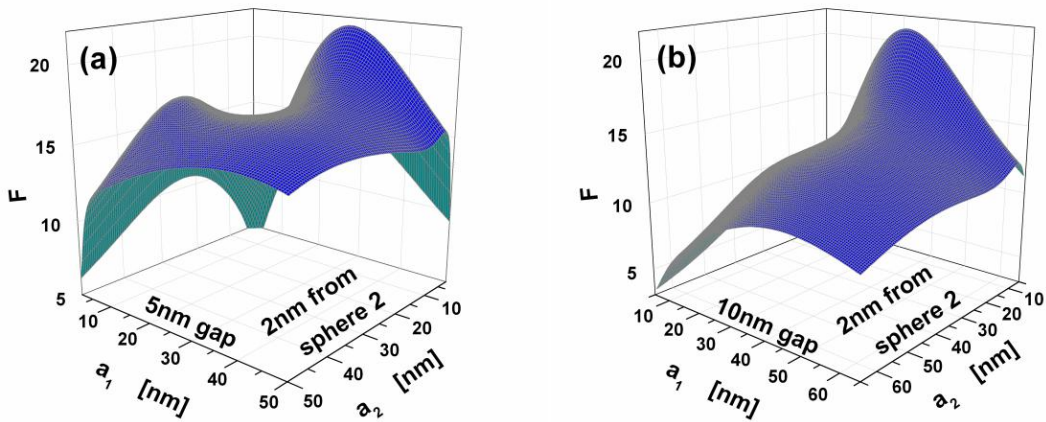

Fig. 4 Enhancement $F$ at 2 nm from sphere 2 in the gap of the two Au spheres embedded in GaN as a function of their radii $a_1$ and $a_2$ with (a) 5 nm and (b) 10 nm gap.

We can now perform optimization of the nanoparticle sizes to obtain peak enhancement $F_{opt}$ at mid gap (Fig. 5(a)) and at 2 nm away from sphere 2 (Fig. 5(b)) for a range of gap sizes. In comparison between the results obtained for the two cases, it can be said that in general the optimal enhancement is somewhat greater for the locations that are closer to one of the spheres than at mid gap for the same gap. The maximum enhancement is on the order of a factor of 20 for the gaps of about 5 nm and approaches 30 for the 2 nm gaps, while Eqs. (20) and (21) predict enhancement as high as almost 300 for $Q\sim 10$ and zero gap. The discrepancy is easily accounted for by the presence of non-zero gap. These results are consistent with those obtained in Ref.[50] which applied the Mie theory to evaluate the enhancement of $E^4$ in the gap of two Ag spheres for the surface enhanced Raman scattering (SERS) process. The enhancement in Ref. [50] is $\sim 10^8$ for a gap of 2nm between two Ag spheres at an off-center location (0.5nm away from one of the spheres), while ours in Fig. 5(a) gives $> 10^6$ for $F^4$ at the mid gap of two Au spheres separated by a 2nm gap. The difference of roughly two orders of magnitude in $F^4$ (about a factor of 3 in $F$) can easily be accounted for by the fact that Au is more lossy than Ag and off-center locations closer to one of the spheres experience more enhancement than mid gap.

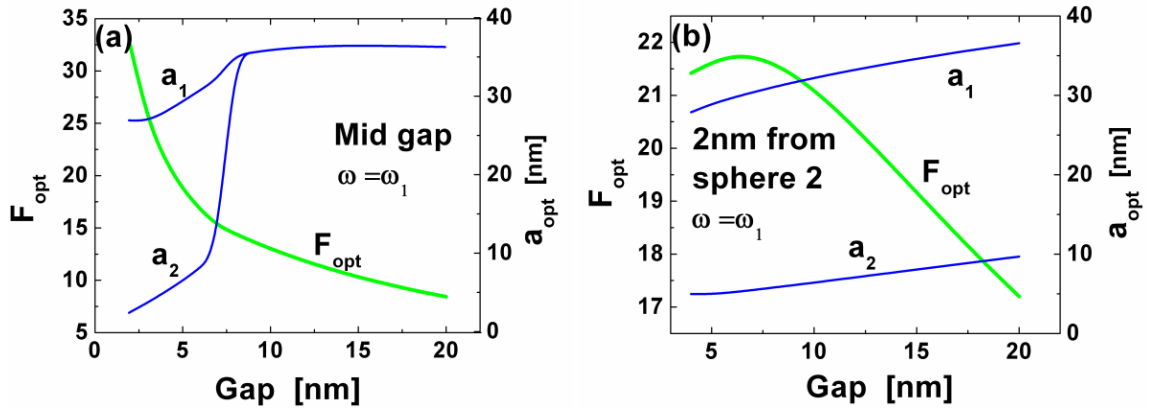

Fig.5 Maximum field enhancement at (a) the mid gap and (b) 2nm away from sphere 2 obtained by optimizing the radii of both spheres as a function of their separation gap.

The above enhancement optimization arrived at the frequency $\omega = \omega_1$ can be further improved by optimizing the frequency. Indeed, the mode coupling can shift the resonance, and in case of strong coupling, it splits into two resonances which can be analyzed by examining the determinant of the $2 \times 2$ matrix $M_{2\times2}$ given by Eq. (18) that is in the denominator of Eq. (19). Two minima at the following two split frequencies from the dipole resonance $\omega_1$ can be obtained

$$\frac{\omega}{\omega_1} \approx 1 \mp \kappa_{11}\left(1 - \frac{1}{4\kappa_{11}^2 Q^2}\right)^{1/2} \qquad (22)$$

when coupling coefficient $\kappa_{11} > 1/2Q$, i.e., the splitting must be greater than the broadening of the dipole mode. The splitting will be further shifted by the coupling with higher order modes. Figure 6 shows the frequency dependence of the enhancement at several gaps with optimized metal sphere radii given in Fig. 5, where the peak enhancement has clearly shifted towards lower frequency $\omega < \omega_1$. The curve for 5 nm gap in Fig. 6(a) and all those in Fig. 6(b) exhibit only one peak with no splitting because their optimized radii are all unequal ($a_1 \neq a_2$) (Fig. 5) which yield small coupling coefficient $\kappa_{11}$ in the range of 0.023~0.038 less than $1/2Q \approx 0.05$. For the frequency dependence of 10 and 20 nm gaps in Fig. 6(a), the sizes of two spheres are optimized at the same radius as shown in Fig. 5 ($a_1 = a_2$), their coupling coefficients $\kappa_{11} = 0.084$ for 10 nm gap and $\kappa_{11} = 0.06$ for 20 nm gap are both greater than $1/2Q$, as a result, a shoulder on the higher frequency side $\omega > \omega_1$ can be resolved revealing the higher split of the dipole resonance. The amount of splitting depends on the coupling strength which is determined by the separation gap – the smaller the gap, the stronger the coupling, and thus the greater the splitting. As the gap increases, the enhancement decreases as a result of the reduced coupling between the spheres.

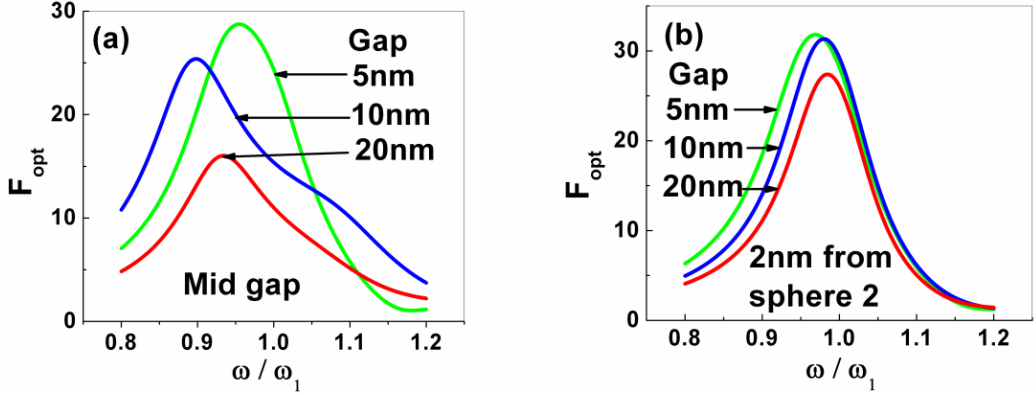

Fig. 6 Frequency dependence of the enhancement at (a) the mid gap and (b) 2nm away from sphere 2 for a range of gap at optimized sphere radii given in Fig.5.

Now let us compare these results with those of single spheres. To have a fair comparison, we obtain optimal enhancement at the locations of equal separation from metal surface for both cases. This means that for a single sphere we are evaluating optimal field enhancement at a separation distance $d$ (normalized distance $\chi_d = 2\pi d/\lambda_1$) which, following our earlier work [60] on a single metal nanosphere, can be obtained at the resonance $\omega = \omega_1$

$$F_{s,opt} = \frac{\sqrt{2}Q}{[1 + (2Q\chi_d^3/3\varepsilon_D)^{1/4}]^4} \tag{23}$$

at the optimized radius $\chi_{opt} = (3\varepsilon_D \chi_d/2Q)^{1/4}$. The result of $F_{s,opt}$ for a single Au sphere embedded in GaN is shown in the insert of Fig. 7(a) for separations up to 10 nm corresponding to 20 nm gap which ensures the entire dimension of two spheres with gap remains smaller than a quarter of the wavelength. The ratio of $F_{opt}/F_{s,opt}$ versus the gap is shown in Fig. 7(a) for mid gap enhancement, and in Fig. 7(b) for the case of 2 nm separation from the sphere. It can be stated that the enhancement in the gaps of coupled spheres always outperforms that of single spheres. The improvement over single sphere is about a factor of 3~4. This factor is substantially smaller than the factor of $2Q \sim 20$ obtained in Eqs. (13) and (15) in the limit of zero gap, but it can be explained by the strong cubic dependence of the field enhancement on the gap width.

For optical absorption and emission with properties directly proportional to the energy density, i.e. electric field squared ($E^2$), the improvement is roughly a factor of 10. For the SERS process whose intensity is proportional to $E^4$, an additional factor of 10 can be recovered. Now, two orders of magnitude is a substantial gain and thus using coupled nanoparticles is definitely worthwhile.

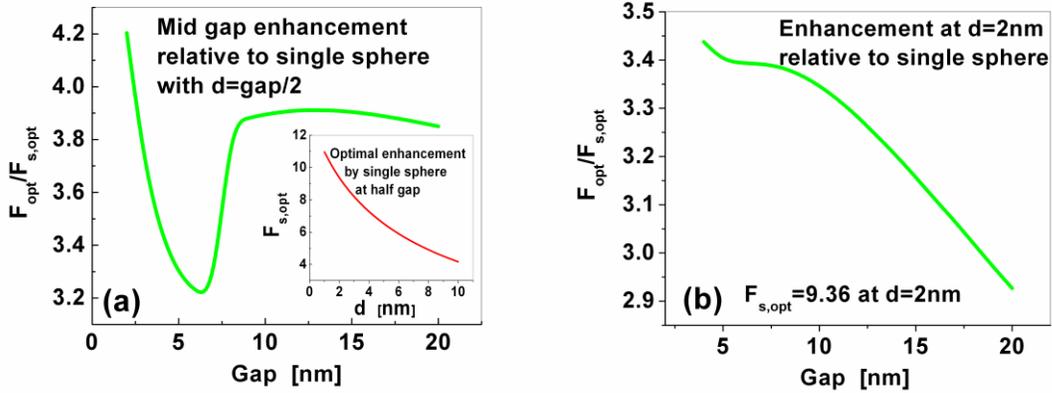

Fig. 7 The ratio of maximum field enhancement by the coupled spheres to that by a single sphere vs. the gap for the case of (a) mid gap (a) and (b) 2 nm separation from one of the spheres. Insert in (a): maximum enhancement by a single sphere vs. the separation $d$ which is equal to half the gap between the two spheres

## VI. Conclusions

In this work we have developed a rigorous analytical approach to the field enhancement in complex systems of coupled metallic nanoparticles. In doing so, we have shown that our previous work [27,57,58,64] can be successfully extended to more complex systems. The main conclusion of our work is the definite evidence that using systems of coupled nanoparticles allows one to achieve larger field enhancements than the ones attainable with a single particle. The simple explanation of this effect is the fact that in order to achieve large enhancement one needs to have both an efficient antenna to interact with incident fields and a small effective mode volume. Single nanopartcles cannot possibly satisfy these two requirements, although a certain degree of optimization is possible. But having more than a single nanoparticle immediately opens up a possibility of using a large dipole mode of one sphere as an efficient antenna and then transfer the energy into one or more tightly confined modes of the other in which the high energy concentration gets achieved. Using an example of two coupled spherical nanopartcles, we have shown that there are two ways the concentration can be achieved. In the case of symmetric dimer, the superposition of quadrupole and higher order modes of both spheres has high energy concentration in the gap between the spheres and this combined "supermode' acts as a small cavity coupled to the dipole antenna. In the case of highly asymmetric "nanolens" the smaller particle acts as a small cavity while the larger particle acts as a dipole antenna. Our theoretical analysis has shown that for both "dimer" and "nanolens" the electric-field enhancement on the order of $Q^2$ near the metal surface can be achieved versus $Q$ in a single particle.

With $Q \approx 10$ for gold, this enhancement would translate into 4 order of magnitude enhancement of the absorption and up to 8 orders of magnitude for luminescence and Raman scattering. This maximum enhancement is reduced, however, once the detuning between different modes, the finite size of the gap, and the distance from the metal surface are taken into account and the optimized field enhancement on the order of 30 appears to be a realistic maximum, which is larger than the enhancement attainable with the single sphere by a factor of about 3~4 and can be translated into about ten-fold improvement for the processes of optical absorption and emission and about 100-fold for SERS.

Compared to numerical calculations, our analytical method clearly offers better physical insight. But the main contribution of our work lies in the fact that this method can be rather easily applied to nano clusters that are far more complex than dimers and trimers to which numerical solutions require complex procedures that are time consuming without a clear strategy for optimization. Using our method, however, one only needs to set up a system of coupled linear equations involving no more than a few modes per sphere, and then perform matrix diagonalization, which is a far less daunting task than a full numerical optimization especially in 3-dimensional case. Furthermore, our method also allows for a quick estimate of the field enhancement achievable at various locations in a complex nano cluster by simply following the progression of energy transfer from optical excitation through mode interactions all the way to the hot spots of interest. Our coupled mode approach thus provides the scientific community with a powerful tool for understanding, estimating, analyzing, and optimizing the metal nanostructures for wide variety of applications.

50. H. Xu and M. Käll, Topics Appl. Phys. 103, 87 (2006).
51. V. Lebedev, S. Vergeles, and P. Vorobev, Opt. Lett. **35**, 640 (2010). The results obtained in this reference depend only on the ratio of the gap and the radius of the nanoparticle and contain no explicit dependence on the material parameters (*Q*). Lack of dependence on the absolute value of the radius indicates that radiative losses are low compared to the Ohmic losses, the case which is true only for very small nanoparticles where the enhancement is actually weak. Furthermore, and far more seriously, the results suggest that the field enhancement for metallic spheres dimer in quasi-static approximation with Drude response formally becomes resonant (infinite) at certain frequency irrespective of the gap size separating them, limited only by losses in the material. However, this contradicts earlier results [52] for this textbook case [53]. The field in the gap becomes formally infinite only for nearly touching spheres irrespective of the frequency in the employed quasistatic approximation [see an exact solution in e.g. Smythe and Klimov mentioned above]. The asymptotic law for the field between nearly touching spheres suggested in the reference is apparently incorrect too.
52. V. V. Klimov and D. V. Guzatov, Phys. Rev. B **75**, 024303 (2007);
53. W. R. Smythe, Static and Dynamic Electricity, 3$^{rd}$ edition (Taylor and Francis, Bristol, 1989), Sec. 5.08
54. A. Kinkhabwala, Z. Yu, S. Fan, Y. Avlasevich, K. Müllen, and W. E. Moerner, Nat. Photonics **3**, 654 (2009).
55. D. R. Ward, F. Hueser, F. Pauly, J. C. Cuevas, and D. Natelson, Nat. Nanotech. **5**, 732 (2010)
56. A. Wokaun, J. P. Gordon P. F. Liao, Phys. Rev. Lett. **48**, 957 (1982).
57. J. B. Khurgin and G. Sun, J. Opt. Soc. Am. B **26**, B83 (2009).
58. J. B. Khurgin, G. Sun, and C. C. Yang, Appl. Phys. Lett. **93**, 1771103 (2009).
59. S. A. Maier, Opt. Express **14**, 1957 (2006).
60. J. B. Khurgin, G. Sun, and R. A. Soref, Appl. Phys. Lett. **93**, 021120 (2008).
61. H. A. Haus, Waves and Fields in Optoelectronics, 1st ed. (Prentice-Hall, Englewood Cliffs, New Jersey, 1984)
62. P. B. Johnson and R. W. Christy, Phys. Rev. B **6**, 4370 (1972).
22